\documentclass[preprint,12pt, sort&compress]{elsarticle}

\usepackage{amssymb}
\usepackage{amsmath}
\usepackage{xcolor}
\usepackage[normalem]{ulem}
\usepackage{lipsum}
\usepackage{array}
\usepackage{soul}
\usepackage{url}
\usepackage{hyperref}

\journal{Solar Energy}

\begin{document}

\begin{frontmatter}



\title{Advancing Lead-Free FASnI$_3$ Perovskite Solar Cells via Cu-Doped NiO$_x$ HTL Engineering: An Optoelectronic Analysis}



\author[inst1]{Md. Faiaad Rahman}
\author[inst1]{Jabed Hasan}
\author[inst1,inst2]{Md. Ashaduzzaman Niloy}
\author[inst1]{Ahmed Zubair\corref{cor1}}
\ead{ahmedzubair@eee.buet.ac.bd}

\cortext[cor1]{Corresponding authors:}

\affiliation[inst1]{organization={Department of Electrical and Electronic Engineering}, 
            addressline={Bangladesh University of Engineering and Technology}, 
            city={Dhaka}, 
            postcode={Dhaka-1205}, 
            country={Bangladesh}}
\affiliation[inst2]{organization={Department of Electrical and Electronic Engineering}, 
            addressline={Green University of Bangladesh}, 
            city={Dhaka}, 
            postcode={Narayanganj-1461}, 
            country={Bangladesh}}
\begin{abstract}
Efficient lead-free perovskite solar cells remain limited by inefficient charge transport and significant recombination losses. In this work, copper (Cu) doped nickel oxide (NiO\textsubscript{x}) is employed as the hole transport layer (HTL) in a formamidinium tin iodide (FASnI\textsubscript{3}) perovskite solar cell to improve hole extraction and suppress carrier losses. The optoelectronic response of the proposed device architecture is evaluated through a coupled numerical framework combining finite-difference time-domain (FDTD) optical simulations with finite element method (FEM)-based electrical modeling. Electrical analysis solves the Poisson equation together with the drift-diffusion and carrier continuity equations, while accounting for radiative, non-radiative, and interfacial recombination mechanisms. Systematic optimization of the thicknesses and doping concentrations of the constituent layers was performed to establish a high-performance device configuration. Following optimization of layer dimensions and carrier densities, the device achieved a power conversion efficiency (PCE) of 24.71\%. Integrating an anti-reflection coating (ARC) effectively suppressed reflection losses, yielding a 5.47\% relative enhancement in PCE. Consequently, the optimized device achieved a peak PCE of 26.06\%, with a short-circuit current density ($J_{\text{sc}}$) of 28.98 $\text{mA/cm}^2$, an open-circuit voltage ($V_{\text{oc}}$) of 1.051 V, and a fill factor (FF) of 85.64\%. These findings demonstrate that $\text{Cu-doped NiO}_x$ functions as a highly promising HTL to boost charge collection and maximize overall efficiency in lead-free $\text{FASnI}_3$ solar cells.

\end{abstract}



\end{frontmatter}
\section{Introduction}
Perovskite solar cells (PSCs) are among the most promising photovoltaic technologies, offering excellent optoelectronic properties and rapid improvements in power conversion efficiency (PCE). Single-junction PSCs now exceed 27\% PCE, and tandem configurations have reached nearly 35\% \cite{green2024solar}. Although significant progress has been achieved, the most advanced PSCs reported to date continue to rely on lead-based absorber materials, raising serious environmental and human health concerns due to lead toxicity and the potential risk of soil and water contamination during large-scale deployment~\cite{torrence2023environmental, dissanayake2023environmental}.
Concerns regarding the toxicity of Pb-based absorbers have prompted extensive research into lead-free perovskite alternatives. Bismuth-, antimony-, and germanium-based materials, together with double perovskites, have been investigated; however, their photovoltaic performance is often constrained by wide bandgaps, limited carrier mobility, or poor defect tolerance.~\cite{byranvand2022tin}. Compared with many other lead-free candidates, tin(Sn) halide perovskites such as MASnI\textsubscript{3}, FASnI\textsubscript{3}, and CsSnI\textsubscript{3} retain several characteristics that are desirable for photovoltaic operation. Sn-based perovskites combine strong visible-light absorption with narrow direct bandgaps and favorable charge transport. These properties make Sn-based perovskites promising absorber materials for efficient lead-free PSCs~\cite{zhu2022smooth}.\\

Among lead-free absorbers, Sn-based perovskites are of particular interest because replacing Pb with Sn can retain many of the electronic characteristics required for photovoltaic operation. MASnI\textsubscript{3}, FASnI\textsubscript{3}, and CsSnI\textsubscript{3} exhibit relatively narrow bandgaps and strong absorption in the visible region. Their charge-transport properties also allow photogenerated carriers to move through the absorber toward the selective contacts. These characteristics have motivated the use of Sn-based compounds as absorber layers in lead-free PSCs. 
 FA/Cs mixing has been reported to improve the structural and thermal stability of Sn-based perovskites compared with single-cation systems \cite{moiz2023lead}. FASnI\textsubscript{3} is therefore widely considered as an absorber because its bandgap of $\sim$1.41 eV is suitable for solar-energy conversion and its thermal stability is higher than that of MASnI\textsubscript{3}~\cite{li2024multi,galve2024addressing}. A major limitation, however, is the rapid oxidation of Sn\textsuperscript{2+}, which promotes defect formation and non-radiative recombination. These losses reduce the attainable V\textsubscript{oc} and adversely affect device stability~\cite{karim2023inhibition,wu2022heterogeneous}.\\

The hole transport layer (HTL) strongly influences charge extraction and interface stability in tin-based PSCs. Organic materials such as PEDOT:PSS, Spiro-OMeTAD, and PTAA are commonly used because their energy levels can support hole extraction. However, these materials are not always well suited to Sn-based devices. Hygroscopicity, chemical instability, dopant-induced degradation, and unfavorable interaction with Sn-perovskites can limit their long-term performance \cite{li2021brief}. These limitations have encouraged the use of inorganic HTLs, including nickel oxide (NiO\textsubscript{x}), doped spinel oxides, and other metal oxides. Compared with many organic HTLs, these materials generally provide better thermal and chemical stability while maintaining good optical transparency \cite{shin2019metal}. Among inorganic HTLs, NiO\textsubscript{x} is widely considered because of its wide bandgap and suitable valence-band position. However, pristine NiO\textsubscript{x} often shows limited p-type conductivity and imperfect energy-level alignment with FASnI\textsubscript{3}, which can hinder hole extraction and promote interfacial recombination \cite{jang2025role}. To address these issues, compositional modification and metal doping have been investigated as effective strategies for improving the conductivity and interfacial charge-transfer behavior of NiO\textsubscript{x}-based HTLs \cite{wu2022efficient,gonzalez2024enhanced}. Copper incorporation into NiO\textsubscript{x} has been reported to improve hole mobility and modify the energy-level alignment at the perovskites/HTL interface. These changes can facilitate hole extraction and reduce interfacial recombination compared with pristine NiO\textsubscript{x} \cite{hosseinzade2025electrochemical}.
These characteristics make Cu:NiO\textsubscript{x} a promising HTL for improving the efficiency and stability of lead-free PSCs.\\

In this work, we designed and systematically evaluated a MgF\textsubscript{2}/ITO/TiO\textsubscript{2}/\\FASnI\textsubscript{3}/Cu:NiO\textsubscript{x}/Au device architecture using experimentally reported material parameters, focusing on the role of Cu-doped NiO\textsubscript{x} as the HTL.MgF\textsubscript{2} served as an anti-reflection coating to improve optical coupling, and TiO\textsubscript{2} acted as the electron-selective contact. 
\begin{table}[!b]
\caption{Electrical simulation parameters of different materials for device simulations}
\centering
\resizebox{\columnwidth}{!}{%
\begin{tabular}{lcccc}
\hline
\textbf{Parameter} & \begin{tabular}[c]{@{}c@{}} \textbf{TiO$_2$}\\ \cite{Bahrami2024oet}\end{tabular} & \begin{tabular}[c]{@{}c@{}} \textbf{FASnI$_3$} \\ \cite{TARA2021111362, rahman2025unveiling} \end{tabular} & \begin{tabular}[c]{@{}c@{}} \textbf{Cu:NiO$_{\rm x}$}\\ \cite{Puja2025nterface}\end{tabular} & \begin{tabular}[c]{@{}c@{}} \textbf{NiO$_{\rm x}$}\\ \cite{kundu2025}, \cite{Wang2018} \end{tabular} \\ \hline
\begin{tabular}[c]{@{}l@{}}Thickness (nm)\end{tabular} & 10--100 & 200--1000 & 20--140 & 20--140 \\ 
Bandgap, $E_g$ (eV) & 3.2 & 1.41 & 3.17 & 3.7 \\ 
\begin{tabular}[c]{@{}l@{}}Relative permittivity, \\$\varepsilon_{r}$\end{tabular} & 9 & 8.2 & 11.7 & 11.7 \\ 
\begin{tabular}[c]{@{}l@{}} Electron affinity,\\$\chi$ (eV) \end{tabular}& 4 & 3.9 & 2.1 & 1.95 \\ 
\begin{tabular}[c]{@{}l@{}}Mobility, $\mu_n/\mu_p$\\(cm$^2$V$^{-1}$s$^{-1}$)\end{tabular} & 20 / 10 & 22 / 22 & 12 / 25 & 10 / 20 \\ 
\begin{tabular}[c]{@{}l@{}} SRH carrier lifetime, \\$\tau_{SRH, e}$/$\tau_{SRH, h}$(ns) \end{tabular}& 5 / 2 & 23.17 & 7.64 & 1.35 \\ 
N\textsubscript{c} (cm$^{\rm-3}$) & $1\times10^{19}$ & $1\times10^{18}$ & $2.8\times10^{21}$ & $1\times10^{19}$ \\ 
N\textsubscript{v} (cm$^{-3}$) & $1\times10^{19}$ & $1\times10^{18}$ & $2.8\times10^{21}$ & $1\times10^{19}$ \\ 
\begin{tabular}[c]{@{}l@{}} Donor density,\\ N\textsubscript{D} (cm$^{-3}$)\end{tabular} & $5\times10^{18}$ & -- & -- & -- \\ 
\begin{tabular}[c]{@{}l@{}} Acceptor density,\\ N\textsubscript{A} (cm$^{-3}$) \end{tabular}& -- & $7\times10^{16}$ & $1.51\times10^{18}$ & $8.1\times10^{18}$ \\ 
\begin{tabular}[c]{@{}l@{}} Surface recombination rate,\\ S (cm/s) \end{tabular}& \begin{tabular}[c]{@{}c@{}} (ITO/TiO$_2$) \\ $1\times10^{7}$ \end{tabular}& -- & \begin{tabular}[c]{@{}c@{}} (Cu:NiO$_{\rm x}$/Au)\\$1\times10^{7}$ \end{tabular}& \begin{tabular}[c]{@{}c@{}}NiO$_{\rm x}$/Au)\\$1\times10^{7}$ \end{tabular} \\ 
\begin{tabular}[c]{@{}l@{}} Radiative recombination\\ coefficient, B\textsubscript{rad} (cm$^3$s$^{-1}$) \end{tabular}& -- & $2.3\times10^{-10}$ & -- & -- \\ 
\begin{tabular}[c]{@{}l@{}} Defect density,\\ N\textsubscript{t} (cm$^{-3}$) \end{tabular}& 1$\times$10\textsuperscript{15} & 2$\times$10\textsuperscript{15} & 10\textsuperscript{15} & 10\textsuperscript{15} \\ \hline
\end{tabular}%
}
\label{tab:material_parameters}
\end{table}
We assessed Cu:NiO\textsubscript{x} in terms of energy-level alignment, hole extraction, and interfacial recombination at the perovskite/HTL interface to enhance the efficiency and operational stability of lead-free FASnI\textsubscript{3}-based solar cells.

\section{Device Modeling and Simulation Methodology}
\begin{figure}[!t]
    \centering
    \includegraphics[width=1.0\linewidth]{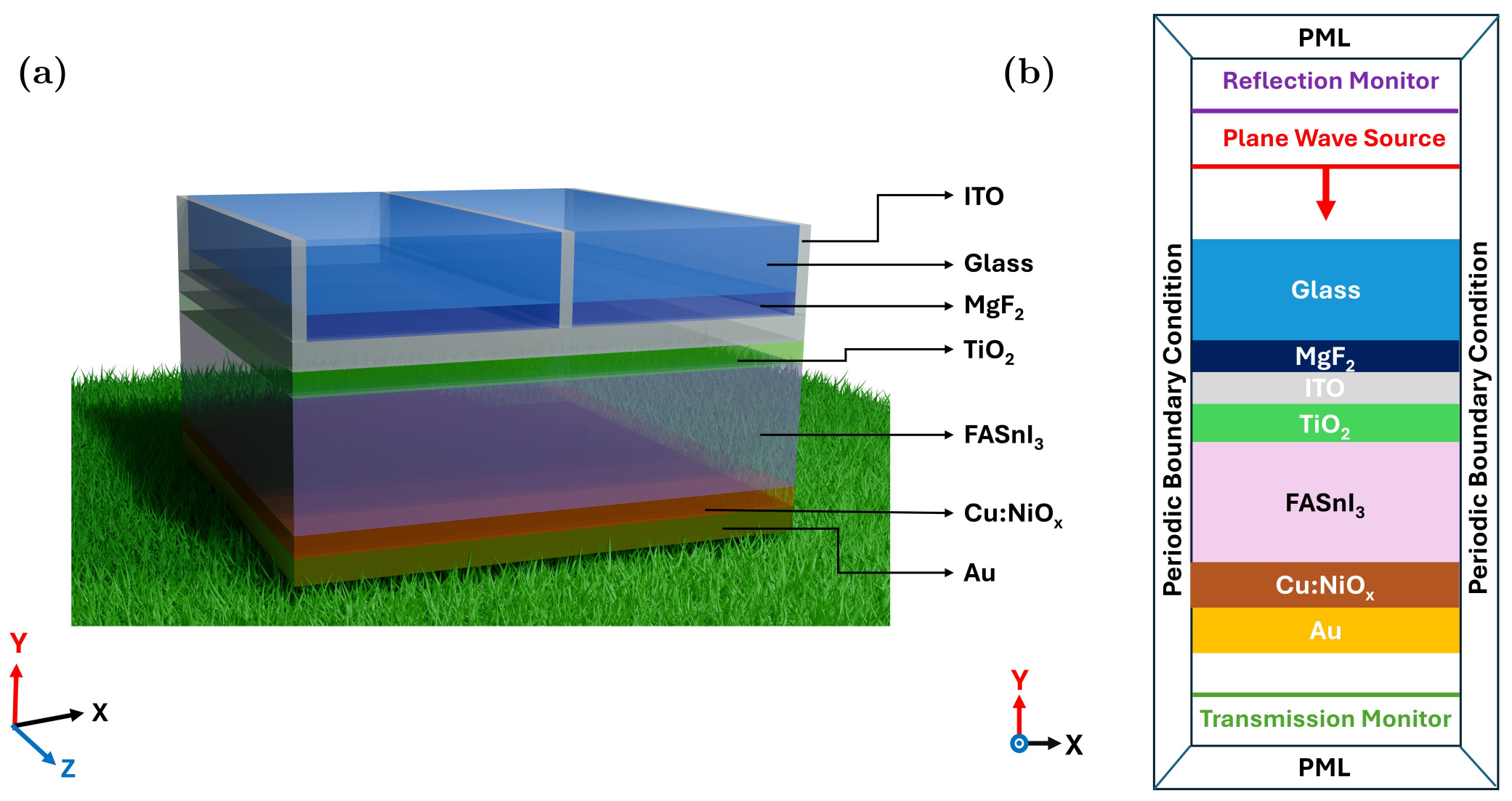}
    \caption{Device architecture and optical simulation setup: 
(a) modeled single-junction solar cell 
(Glass/MgF$_2$/ITO/TiO$_2$/FASnI$_3$/Cu:NiO$_x$/Au) and 
(b) two-dimensional FDTD boundary conditions for optical analysis.}
    \label{fig1}
\end{figure}

The proposed single-junction solar cell was modeled in an n+/p/p+ configuration, as illustrated in Fig.~\ref{fig1}(a). The photoactive region consists of p-type doped FASnI\textsubscript{3}, a direct-bandgap perovskite semiconductor with a bandgap of 1.41 eV, which serves as the primary light-absorbing layer. A compact n-type doped titanium dioxide (TiO\textsubscript{2}) layer, having a wide bandgap of 3.2 eV, was employed as the electron transport layer (ETL) to facilitate selective electron extraction. Owing to its favorable hole-transport characteristics and suitable interfacial properties, p-type doped copper-doped nickel oxide (Cu:NiO\textsubscript{x}) was incorporated as the hole transport layer (HTL). Magnesium fluoride (MgF\textsubscript{2}) was introduced as an anti-reflection coating to minimize optical reflection losses, whereas indium tin oxide (ITO) functioned as the transparent conducting oxide (TCO) and front electrode. Finally, gold (Au) was employed as the back contact layer (BCL) to complete the device architecture.\newline

\hspace*{1em} The optoelectronic response of the proposed device was evaluated using the Ansys Lumerical FDTD and CHARGE solvers. Fig.~\ref{fig1}(b) illustrates the boundary conditions employed in the two-dimensional FDTD model for optical analysis. Periodic boundary conditions were imposed along the X-direction to represent the lateral periodicity of the planar device, while perfectly matched layer (PML) boundary conditions were applied along the Y-direction to suppress artificial reflections from the simulation boundaries. The device was illuminated from the top along the Y-direction using the standard AM1.5G solar spectrum over a wavelength range of 200 to 1200 nm. The wavelength-dependent complex refractive indices of the constituent materials were used as optical inputs to calculate the spatial distribution of optical absorption and photogenerated carriers throughout the device structure. The refractive-index data for MgF$_2$, ITO, TiO$_2$, FASnI$_3$, Cu:NiO$_x$, and Au were adopted from previously reported literature~\cite{Rodriguez-deMarcos:17,Holman2013ITO,Sarkar2019TiO,Ghimire2017FSI,Manzoor:18NiO,Magnozzi2019Au}. The resulting carrier generation profile obtained under AM1.5G illumination was subsequently imported into the CHARGE solver to perform the coupled electrical analysis and evaluate the overall optoelectronic performance of the device.\\

For the electrical analysis, carrier transport within the device was evaluated using the CHARGE solver by self-consistently solving the coupled Poisson and carrier continuity equations under appropriate Neumann and Dirichlet boundary conditions. To provide a realistic representation of carrier loss mechanisms, the model incorporated radiative, non-radiative, and surface recombination processes throughout the device.
The electrical properties assigned to each constituent material were selected from experimentally validated values reported in recent literature, as summarized in Table \ref{tab:material_parameters}.
A sequential optimization procedure was used to determine the device configuration. The FASnI\textsubscript{3} absorber thickness was varied first, followed by the ETL and HTL thicknesses. The doping concentrations of the HTL, ETL, and absorber were then adjusted sequentially while keeping the previously optimized parameters fixed. 
In the final stage, the TCO and ARC thicknesses were refined to minimize optical losses and improve photon coupling within the device structure. The photovoltaic performance of each configuration was evaluated using the power conversion efficiency (PCE), open-circuit voltage (V\textsubscript{oc}), short-circuit current density (J\textsubscript{sc}), and fill factor (FF), extracted from the simulated current--voltage characteristics.

\section{Results and Discussion}
\begin{figure}[!t]
    \centering
    \includegraphics[width=1.0\linewidth]{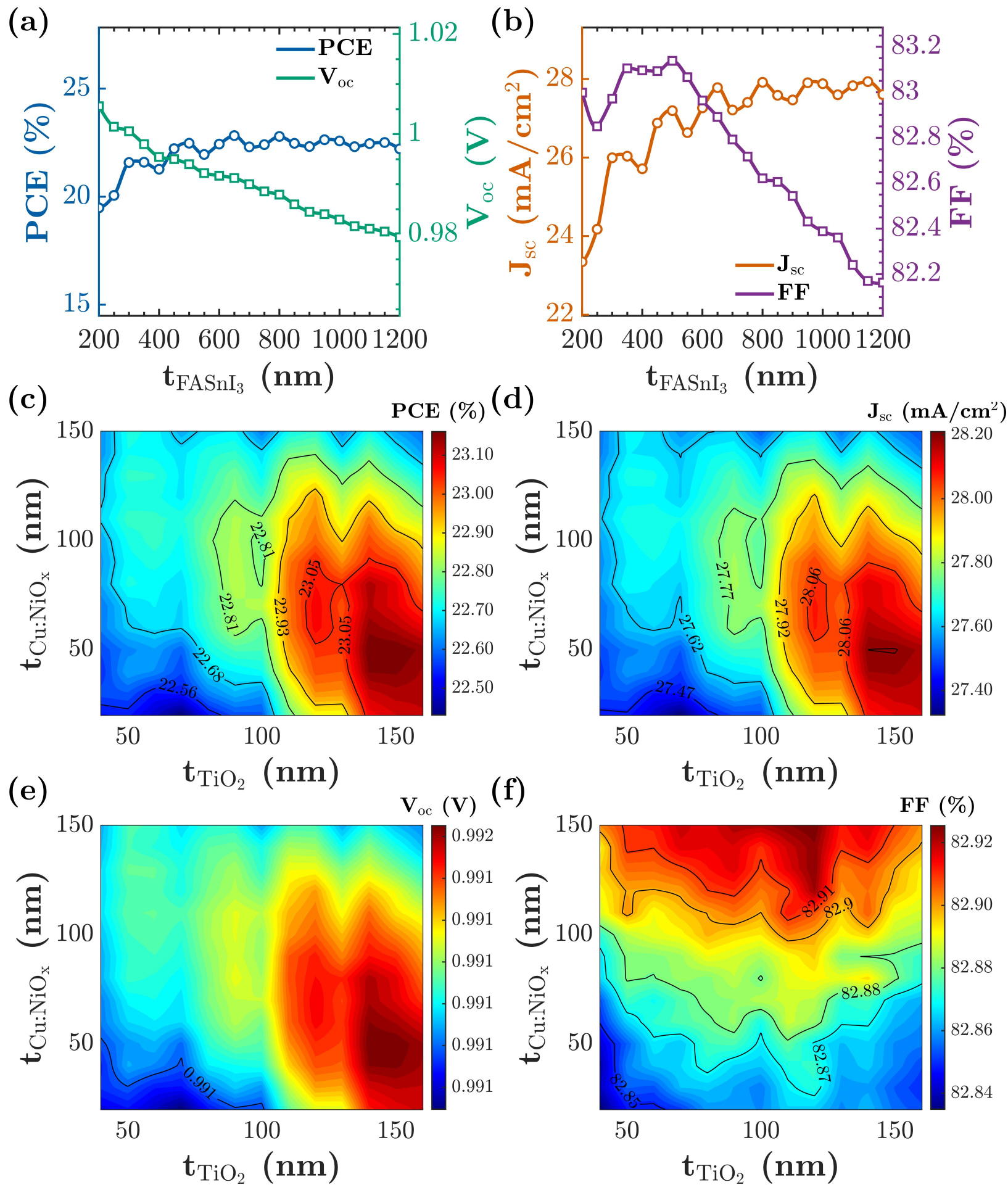}
    \caption{Impact of absorber thickness variation on the photovoltaic performance metrics: (a) PCE ($\eta$) \& V$_{oc}$ (b)J$_{sc}$ \& FF. Impact of TiO$_2$ ETL thickness variation on the photovoltaic performance metrics: (c) PCE ($\eta$), (d) J$_{sc}$, (e) V$_{oc}$, and (f) FF.}
    
    \label{fig2}
\end{figure}
Following the established simulation framework, a baseline device configuration was first constructed using ITO, TiO\textsubscript{2}, FASnI\textsubscript{3}, Cu:NiO\textsubscript{x}, and Au layer thicknesses of 100 nm, 100 nm, 200 nm, 100 nm, and 100 nm, respectively. The TiO\textsubscript{2} ETL was modeled as an n-type semiconductor with a donor concentration (N\textsubscript{D}) of 5$\times$10\textsuperscript{18} cm\textsuperscript{-3}, while the FASnI\textsubscript{3} absorber and Cu:NiO\textsubscript{x} HTL were considered p-type with an acceptor concentration of 1.51$\times$10\textsuperscript{18} cm\textsuperscript{-3}. Under these initial conditions, the reference device exhibited a PCE of 19.49\%, with a J\textsubscript{sc} of 23.35 mA/cm\textsuperscript{2}, a V\textsubscript{oc} of 1.01 V, and an FF of 82.99\%. This baseline configuration was subsequently used as the reference for the sequential optimization of the device parameters.

\subsection{Optimization of layer thicknesses and doping densities}
The thicknesses of the functional layers were systematically optimized to balance optical absorption, carrier collection, and recombination losses. The FASnI$3$ absorber thickness was varied from 200 to 1200 nm. As shown in Fig.~\ref{fig2}(a), the PCE increased markedly from approximately 19.5\% at 200 nm and reached a maximum of 22.82\% at 650 nm, after which it exhibited only small fluctuations followed by a slight overall reduction. The corresponding photovoltaic parameters in Fig.~\ref{fig2}(b) clarify this behaviour. Increasing the absorber thickness enhanced photon absorption and photocarrier generation, causing J$_{\mathrm{sc}}$ to increase from approximately 23.3 to nearly 28 mA/cm$^2$. In contrast, V$_{\mathrm{oc}}$ decreased progressively with thickness, while the FF also showed an overall reduction at larger thicknesses. These reductions are attributed to the increased carrier diffusion length and enhanced bulk recombination in the thicker absorber, which increase carrier losses and limit the gain obtained from additional optical absorption. Consequently, the optimum thickness of 650 nm represents the best compromise between the increasing J$_{\mathrm{sc}}$ and the concurrent reductions in V$_{\mathrm{oc}}$ and FF, resulting in the maximum PCE.\\

\begin{figure}[!t]
    \centering
    \includegraphics[width=1.0\linewidth]{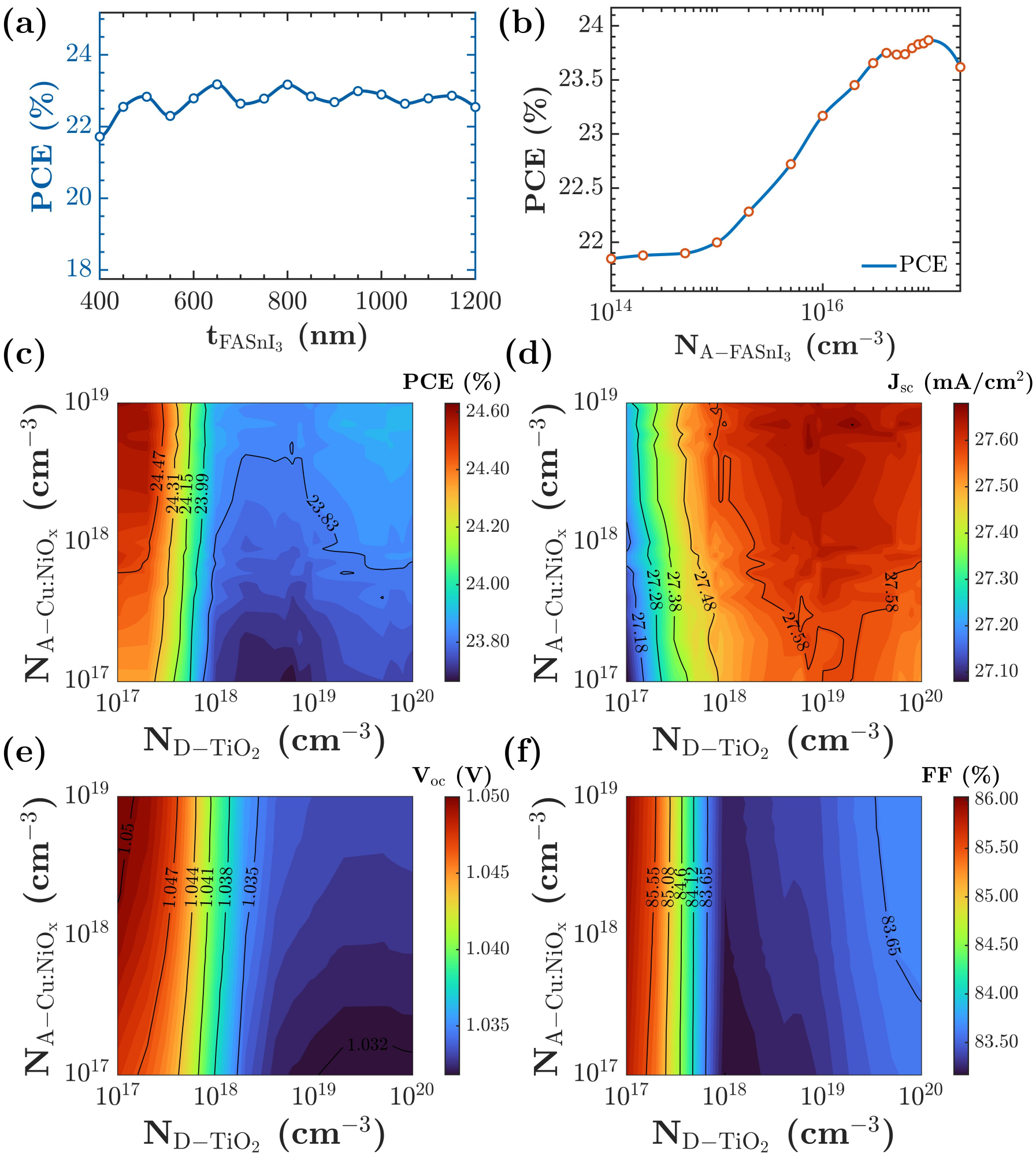}
    \caption{Optimization of the FASnI$_3$-based device: PCE ($\eta$) variation with (a) absorber thickness and (b) absorber doping, and the effects of TiO$_2$/Cu:NiO$_x$ doping on (c) PCE, (d) J$_{sc}$, (e) V$_{oc}$, and (f) FF.}
    \label{fig:3}
\end{figure}

Figs.~\ref{fig2}(c-f) present the contour maps of PCE, J\textsubscript{sc}, V\textsubscript{oc}, and FF as functions of ETL and HTL thickness in the 40--160~nm range and 20--180~nm range, respectively. The PCE trends as shown in Fig.~\ref{fig2}(a) are primarily governed by charge extraction at the absorber/transport-layer interfaces rather than bulk absorption, as indicated by the weak sensitivity of V\textsubscript{oc} and FF across the sweep featured in Figs.~\ref{fig2}(b) and \ref{fig2}(c). For thin ETL layers ($<$30 nm), incomplete electron selectivity and increased interfacial recombination reduce J\textsubscript{sc}. Increasing the ETL thickness in the range of ~80--100 nm strengthens band bending and reduces interface recombination, leading to higher current density. A similar behavior is observed for the HTL, where thin Cu:NiOx layers inadequately block electrons and limit hole extraction, while thicknesses in the range of ~60–100 nm provide sufficient field screening and quasi-Fermi level splitting, slightly improving V\textsubscript{oc}. Beyond these ranges, performance saturates or marginally decreases due to increased carrier transit length and series resistance. The nearly uniform FF contours in Fig.~\ref{fig2}(f) confirm that the device operates in a recombination-controlled regime with minimal resistive losses.\\

To verify the consistency of the absorber thickness optimization, an additional thickness variation of the FASnI$_3$ layer was performed while maintaining the optimized ETL and HTL thicknesses, as shown in Fig.~\ref{fig:3}(a). The variation exhibited a similar trend, with the optimum absorber thickness remaining at 650~nm, confirming the reliability of the obtained thickness value. After fixing the optimized layer thicknesses, the Cu:NiO\textsubscript{x} HTL acceptor concentration (N\textsubscript{A}) was varied to assess its influence on device performance. As shown in Fig.~\ref{fig:3}(b), increasing N\textsubscript{A} slightly improved the PCE by enhancing hole conductivity and reducing resistive losses. Although a maximum PCE of 24.60\% was obtained at 7$\times$10\textsuperscript{18} cm\textsuperscript{-3}, the improvement beyond 2$\times$10\textsuperscript{18} cm\textsuperscript{-3} was marginal. Therefore, 2$\times$10\textsuperscript{18} cm\textsuperscript{-3} was selected as the optimum trade-off, yielding a PCE of approximately 24.56\%.\newline

Subsequently, the donor doping density, N\textsubscript{D} of the TiO\textsubscript{2} ETL was varied in the same range as the HTL, as noticed in Fig.~\ref{fig:3}(c). At low doping levels, the PCE remains slightly lower due to limited electron conductivity and higher transport resistance in the ETL. Increasing N\textsubscript{D} of ETL enhances electron extraction and reduces resistive losses, leading to a gradual improvement in PCE and reaching a maximum value of $\sim$24.60\% at 2$\times$10\textsuperscript{17} cm\textsuperscript{-3}. Beyond this concentration, the PCE gradually declines as increased carrier density induces higher interface recombination and FF degradation, indicating that excessively high ETL doping is detrimental to device performance. Following that, we investigated the impact of FASnI\textsubscript{3} doping on device performance. As shown in Fig.~\ref{fig:3}(b), N\textsubscript{A} of the FASnI\textsubscript{3} absorber from 1$\times$10\textsuperscript{14} to 2$\times$10\textsuperscript{17} cm\textsuperscript{-3} to evaluate its impact on device efficiency. At low doping levels, the PCE remains limited due to weak built-in electric fields, resulting in inefficient drift-assisted charge separation and reduced quasi-Fermi level splitting. As the absorber doping increases, the internal field strengthens, improving carrier collection and reducing recombination losses, which leads to a steady rise in PCE. The PCE reaches a maximum value of 23.82\% at 8$\times$10\textsuperscript{16} cm\textsuperscript{-3}, where drift-diffusion transport is most effective. Further increase in doping causes efficiency degradation due to enhanced bulk recombination and reduced carrier diffusion length, leading to a decline in PCE.\\

\begin{figure}[!t]
    \centering
    \includegraphics[width=1.0\linewidth]{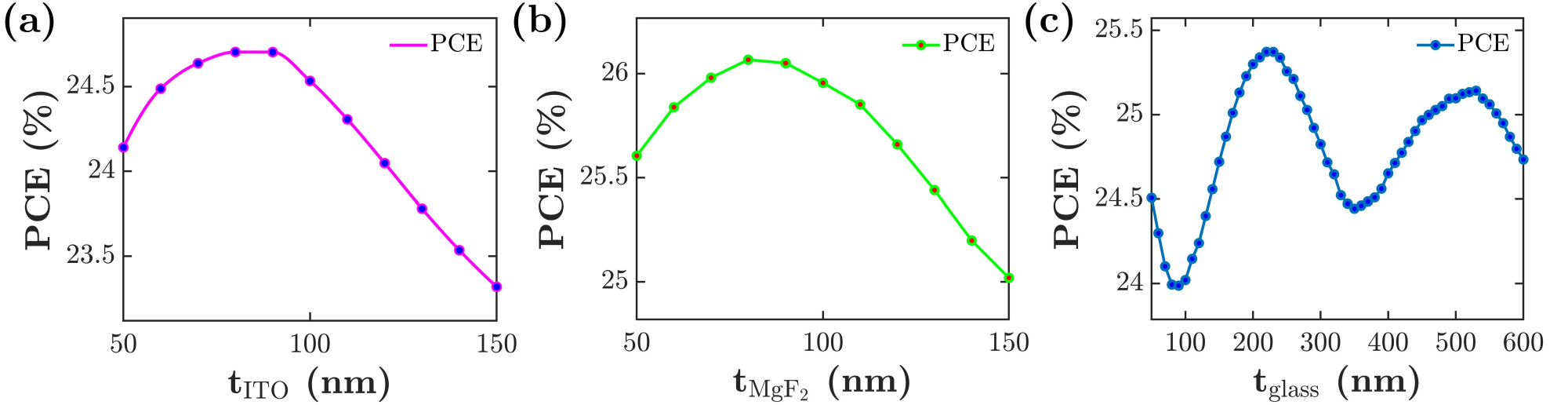}
    \caption{Impact on PCE ($\eta$) corresponding to the thickness variation of (a) ITO, (b) MgF$_2$ ARC, and (c) glass substrate.}
    \label{fig:4}
\end{figure}
\hspace*{1em}Next, the thicknesses of ITO TCO layer and the MgF\textsubscript{2} ARC were optimized. The ITO thickness was varied from 50~nm to 150~nm, gaining an optimal PCE of 24.70\% at an ITO thickness of 80~nm, where electrical conductivity is sufficiently high while optical losses remain limited as seen in Fig.~\ref{fig:4}(a). Concurrently, the MgF$_2$ ARC thickness was varied in the 50--150~nm range to minimize front-surface optical losses. The maximum efficiency was obtained at an ARC thickness of 80~nm. 
As observed from Fig.~\ref{fig:4}(b), the incorporation of the ARC significantly reduces front-surface optical reflection in the 410--550 nm wavelength range, thereby enhancing photon injection into the underlying active layers and consequently increasing the photogenerated carrier density. As shown in Fig.~\ref{fig:4}(c), the PCE exhibits an oscillatory dependence on the glass thickness due to thickness-dependent optical interference and reflection within the multilayer structure. An optimum glass thickness of approximately 220 nm yields the highest PCE, while further variation in thickness produces alternating enhancement and reduction in device efficiency.
A champion PCE of 26.06\%  with J\textsubscript{sc} of 28.98 mA/cm\textsuperscript{2}, V\textsubscript{oc} of 1.051V, and FF of 85.64\% was secured for the optimum structure, surging a relative gain of 5.47\% compared to the model without ARC counterparts.

\begin{figure}[!ht]
    \centering
    \includegraphics[width=0.92\linewidth]{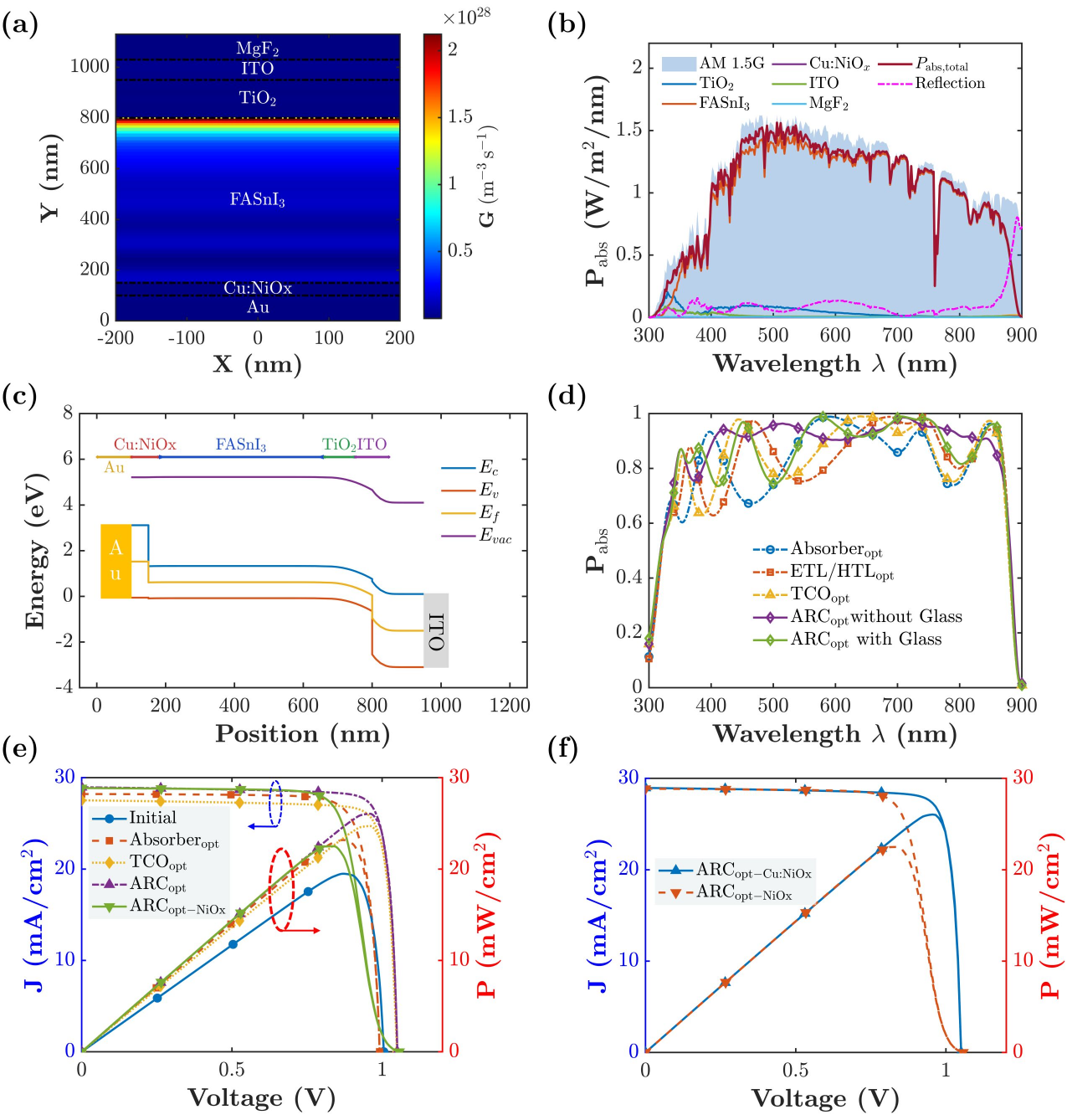}
    \caption{Optoelectronic characteristics of the optimized device: (a) carrier generation rate profile, (b) spectral absorption, (c) energy-band diagram, (d) normalized absorption, and J--V and P--V characteristics: (e) sequential optimization and (f) Cu:NiO$_x$ and NiO$_x$ HTLs.}
    
    \label{fig:5}
\end{figure}
\subsection{Impact of Cu-doped NiOx as HTL}
Cu incorporation improves the p-type conductivity, hole concentration, and carrier mobility of NiO$_x$, while providing more favorable valence band alignment with FASnI$_3$, as shown in Fig.~\ref{fig:5}(c). This improved interfacial alignment facilitates hole extraction, suppresses recombination losses, and enhances charge transport. Consequently, Cu:NiO$_x$ delivers superior photovoltaic performance compared with pristine NiO$_x$, confirming its suitability as an effective HTL for lead-free FASnI$_3$-based solar cells.
\subsection{Progressive performance evolution during optimization}
\begin{table}[!b]
\caption{Summary of the performance of the overall device architecture following different optimization stages.}
\centering
\resizebox{0.6\textwidth}{!}{%
\begin{tabular}{ccccc}
\hline
\begin{tabular}[c]{@{}c@{}}\textbf{Parameters}\end{tabular} & \begin{tabular}[c]{@{}c@{}}$\mathbf{PCE}$ \\(\%)\end{tabular} & \begin{tabular}[c]{@{}c@{}} $\mathbf{J_{sc}}$ \\(mA/cm$^2$) \end{tabular}& \begin{tabular}[c]{@{}c@{}}$\mathbf{V_{oc}}$ \\(V)\end{tabular} & \begin{tabular}[c]{@{}c@{}} $\mathbf{FF}$\\(\%)\end{tabular} \\ \hline
\begin{tabular}[c]{@{}c@{}} Initial Structure \end{tabular} & 19.49 & 23.35 & 1.01 & 82.99 \\ 
\begin{tabular}[c]{@{}c@{}} Thickness optimization \\(ETL/ABS/HTL) \\ (150 nm/ 650 nm/ 50 nm)\end{tabular} & 23.18 & 28.21 & 1.01 & 82.86 \\ 
\begin{tabular}[c]{@{}c@{}} Doping optimization \\ (ETL/HTL) \\($2\times10^{17}$, $2\times10^{18}$ cm$^{-3}$) \end{tabular}& 24.55 & 27.38 & 1.04 & 85.51 \\
\begin{tabular}[c]{@{}c@{}} Doping optimization \\ (Absorber = $8\times10^{16}$ cm$^{-3}$) \end{tabular}& 24.56 & 27.38 & 1.05 & 85.51 \\ 
\begin{tabular}[c]{@{}c@{}} ARC thickness \\(80 nm) \end{tabular}& 26.06 & 28.98 & 1.05 & 85.64 \\ 
\begin{tabular}[c]{@{}c@{}} Incorporating NiO$_{\rm x}$ as HTL\\(with glass) \end{tabular}& 22.49 & 28.86 & 1.056 & 73.74 \\ \hline
\end{tabular}%
}
\label{tab:pv_performance}
\end{table}

Fig.~\ref{fig:5}(a) illustrates the spatial carrier-generation profile, showing that the strongest photogeneration occurs within the FASnI\textsubscript{3} absorber, while Fig.~\ref{fig:5}(b) presents the wavelength-dependent absorption of the individual device layers under the standard AM1.5G spectrum. The energy-band profile shown in Fig.~\ref{fig:5}(c) illustrates the carrier-selective band alignment across the device structure. Fig.~\ref{fig:5}(d) further compares the normalized absorption response at different optimization stages, demonstrating the progressive improvement in photon harvesting.
The corresponding J--V and P--V characteristics in Fig.~\ref{fig:5}(e), together with the photovoltaic parameters summarized in Table~\ref{tab:pv_performance}, demonstrate the sequential enhancement in device performance. The initial configuration exhibited a PCE of 19.49\%. Optimization of the ETL, FASnI\textsubscript{3} absorber, and HTL thicknesses to 150~nm, 650~nm, and 50~nm, respectively, increased the PCE to 23.18\%. Subsequent optimization of the ETL and HTL doping concentrations raised the PCE to 24.55\%, while optimization of the absorber doping concentration further improved it to 24.56\%. Finally, incorporating an 80~nm MgF$_2$ anti-reflection coating reduced optical losses and enhanced photon coupling, resulting in a maximum PCE of 26.06\%. The FASnI\textsubscript{3} absorber exhibits strong photon harvesting across the visible and near-infrared (NIR) spectral regions. The J--V and P--V comparison in Fig.~\ref{fig:5}(f) further highlights the influence of HTL engineering. The device incorporating pristine NiO$_x$ exhibited a considerably lower PCE of 22.49\%, whereas the Cu:NiO$_x$-based configuration achieved superior photovoltaic performance. Cu incorporation improves hole transport and facilitates more favorable carrier extraction at the FASnI$_3$/HTL interface, thereby reducing recombination losses. These results demonstrate the effectiveness of Cu:NiO$_x$ as a hole-selective layer for high-performance lead-free FASnI$_3$ PSCs.
\begin{figure}[!t]
    \centering
    \includegraphics[width=1.0\linewidth]{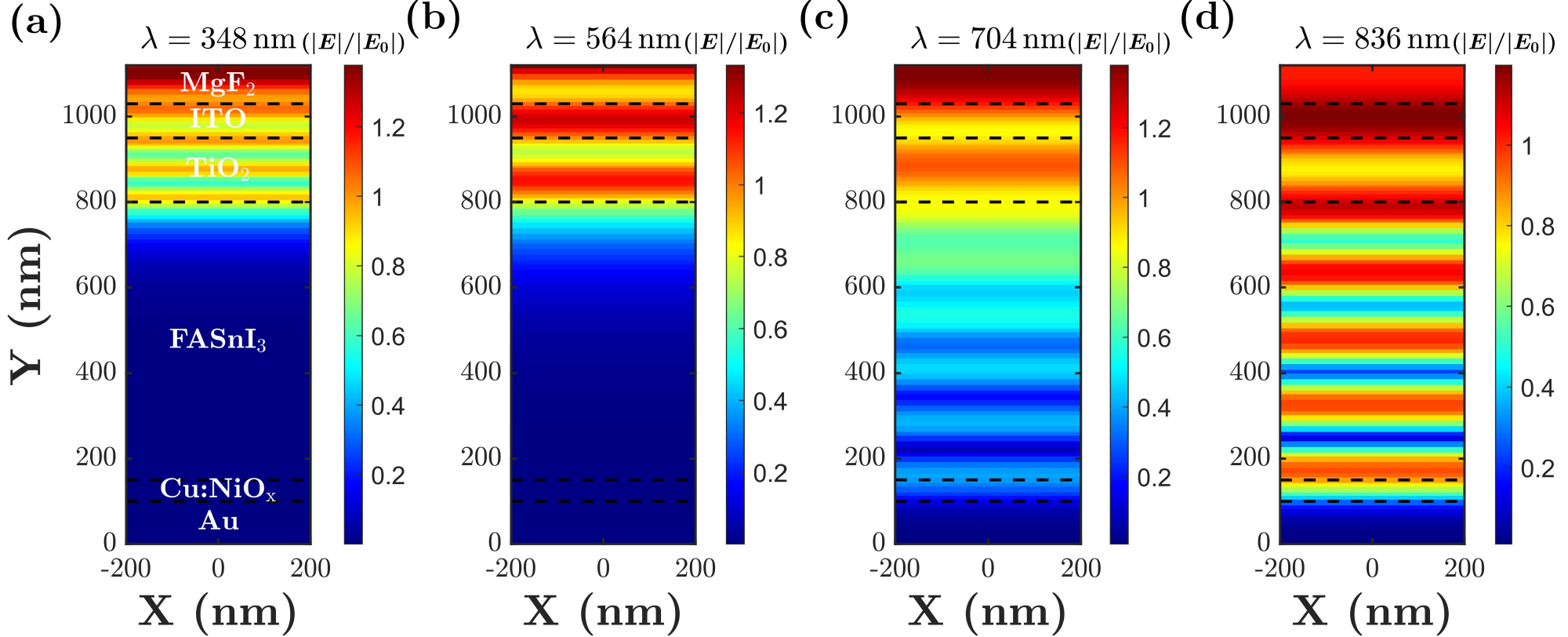}
    \caption{Normalized optical electric-field distribution ($|E|/|E_0|$) within the optimized solar-cell structure at representative wavelengths of (a) 348 nm, (b) 564 nm, (c) 704 nm, and (d) 836 nm. The distributions illustrate the wavelength-dependent field penetration and localization across the constituent layers, particularly within the FASnI$_3$ absorber.}
    \label{fig:6}
\end{figure}

\subsection{Optical Electric-Field Distribution in the Optimized Structure}
For the optimized device, Fig.~\ref{fig:6}(a-d) presents the normalized electric-field distribution, \((|E|/|E_0|)\), at selected wavelengths. At 348 nm and 564 nm, the optical field is the strongest close to the front layers of the device, where it primarily penetrates into the upper part of the FASnI$_3$ absorber. The field intensity drops off quickly as you move deeper into the absorber at these shorter wavelengths. In contrast, when the wavelength increases to 704 nm and 836 nm, the field extends much further into the FASnI$_3$ layer. At these longer wavelengths, you can also observe well-defined spatial oscillations in the field, which result from the interference between the incoming light and the light reflected from the gold (Au) back contact. This interference effect creates a standing-wave pattern within the absorber, leading to alternating regions of higher and lower electric field. Especially at 836 nm, multiple field maxima are seen throughout the FASnI$_3$ layer. Overall, this optimized structure effectively confines the optical field over a broad range of wavelengths, with strong near-surface interaction at shorter wavelengths and enhanced penetration and localization at longer wavelengths. These features improve the chances for weakly absorbed photons to interact with the active layer, which can boost the device's overall efficiency.

\subsection{Comparative analysis with existing literature}

Table \ref{tab:comparison_current_previous} compares the performance of the proposed lead-free Sn-based perovskite solar cell employing Cu:NiO\textsubscript{x} as the HTL with recently reported devices. The results from previous studies are included to benchmark the photovoltaic performance of the present structure. The Cu:NiO\textsubscript{x}-based device demonstrates superior performance to those using PEDOT:PSS and pristine NiO\textsubscript{x}, while remaining comparable to CuSCN-based configurations. The improvement achieved through Cu incorporation highlights the effectiveness of Cu:NiO\textsubscript{x} as a competitive hole transport material for high-performance lead-free perovskite solar cells.

\begin{table}[!htbp]
\centering
\caption{Performance comparison of FASnI$_3$ single-junction solar cells with different ETL/Abs/HTL configurations.}
\resizebox{\columnwidth}{!}{%
\begin{tabular}{cccccc}
\hline
\textbf{\begin{tabular}[c]{@{}c@{}}ETL/Abs/HTL\end{tabular}} 
& \textbf{\begin{tabular}[c]{@{}c@{}}PCE\\(\%)\end{tabular}} 
& \textbf{\begin{tabular}[c]{@{}c@{}}J$_{\rm sc}$\\(mA/cm$^{2}$)\end{tabular}} 
& \textbf{\begin{tabular}[c]{@{}c@{}}V$_{\rm oc}$\\(V)\end{tabular}} 
& \textbf{\begin{tabular}[c]{@{}c@{}}FF\\(\%)\end{tabular}} 
& \textbf{Ref.} \\
\hline
\begin{tabular}[c]{@{}c@{}}Zn(O$_{0.3}$, S$_{0.7}$)/FASnI$_3$/CuSCN\end{tabular} & 25.94 & 28.12 & 1.086 & 84.96 & \cite{TARA2021111362} \\ 
TiO$_2$/FASnI$_3$/PTTA & 27.04 & 30.84 & 1.100 & 79.71 & \cite{Verma2025} \\ 
\begin{tabular}[c]{@{}c@{}}BCP/C$_{60}$/FASnI$_3$/PEDOT:PSS\end{tabular} & 10.16 & 21.95 & 0.640 & 72.50 & \cite{Meng2020} \\ 
\begin{tabular}[c]{@{}c@{}}BCP/ICBA/FASnI$_3$/PEDOT:PSS\end{tabular} & 14.06 & 20.05 & 0.911 & 75.69 & \cite{zhu2022smooth} \\ 
TiO$_2$/FASnI$_3$/CuSCN & 26.46 & 28.28 & 1.110 & 84.02 & \cite{Alqurashi2024} \\ 
TiO$_2$/FASnI$_3$/NiO$_x$ & 22.49 & 28.86 & 1.056 & 73.74 & Our Work \\ 
TiO$_2$/FASnI$_3$/Cu:NiO$_x$ & 26.06 & 28.98 & 1.051 & 85.64 & Our Work \\
\hline
\end{tabular}%
}
\label{tab:comparison_current_previous}
\end{table}

\section{Conclusion}
Achieving high-performance lead-free tin-based perovskite solar cells requires coordinated optimization of interfacial energetics, charge-carrier transport, and optical management. In this study, a lead-free FASnI$_3$ perovskite solar cell incorporating Cu-doped NiO$_x$ as an inorganic hole transport layer was systematically investigated within the MgF$_2$/\allowbreak ITO/\allowbreak TiO$_2$/\allowbreak FASnI$_3$/\allowbreak Cu:NiO$_x$/\allowbreak Au device architecture. The photovoltaic performance was progressively improved through sequential optimization of the functional-layer thicknesses, doping concentrations, and front-surface anti-reflection coating.Optimization of the ETL, absorber, and HTL thicknesses, together with appropriate carrier doping, strengthened the internal electric field, enhanced charge transport, reduced recombination losses, and improved photocarrier collection. The subsequent incorporation of an MgF$_2$ ARC further suppressed optical reflection and enhanced photon harvesting, resulting in a maximum PCE of 26.06\%. In addition, Cu incorporation into NiO$_x$ improved the valence-band alignment at the FASnI$_3$/HTL interface, thereby facilitating hole extraction and reducing non-radiative recombination compared with pristine NiO$_x$.These findings demonstrate that Cu:NiO$_x$ is a promising hole-selective material for efficient lead-free FASnI$_3$ PSCs and highlight the importance of simultaneous optical and electrical optimization in maximizing device performance. The proposed single-junction architecture also provides a potential platform for future integration into multi-junction photovoltaic configurations to further enhance solar-energy conversion efficiency.

\section{CRediT authorship contribution statement}
\textbf{Md. Faiaad Rahman:} conceptualization, methodology, data curation, visualization, software, investigation, writing original draft, and writing review \& editing.
\textbf{Jabed Hasan:} conceptualization, methodology, data curation, visualization, software, investigation, writing original draft, and writing review \& editing.
\textbf{Md. Ashaduzzaman Niloy:} conceptualization, methodology, data curation, visualization, investigation, writing original draft, and writing review \& editing.
\textbf{Ahmed Zubair:} conceptualization, methodology, data curation, visualization, investigation, supervison, writing original draft, and writing review \& editing.
\section{Data availability statement}
The data that support the findings of this study are available from the corresponding author upon reasonable request.
\section{Declaration of competing interest}
The authors declare that they have no known competing financial interests or
personal relationships that could have appeared to influence the work reported in
this paper.
\section{Acknowledgments}
The authors express their sincere gratitude to the Department of Electrical and Electronic Engineering at Bangladesh University of Engineering and Technology (BUET) for providing access to the Ansys Lumerical software and the necessary technical support.




\bibliographystyle{elsarticle-num} 
\bibliography{mybibfile}






\end{document}